# Imaging proteins at the single molecule level

**Authors:** Jean-Nicolas Longchamp[1,*], Stephan Rauschenbach[2], Sabine Abb[2], Conrad Escher[1], Tatiana Latychevskaia[1], Klaus Kern[2,3], Hans-Werner Fink[1]

***Affiliations:***

[1]*Physics Department of the University of Zurich, Winterthurerstrasse 190, CH-8057 Zürich, Switzerland*

[2]*Max Planck Institute for Solid State Research, Heisenbergstrasse 1, DE-70569 Stuttgart, Germany*

[3]*Institut de Physique de la Matière Condensée, Ecole Polytechnique Fédérale de Lausanne, CH-1015 Lausanne, Switzerland*

*\* Corresponding author: longchamp@physik.uzh.ch*

**Classification: Physical Sciences/Applied Physical Sciences**

**Abstract**

Imaging single proteins has been a long-standing ambition for advancing various fields in natural science, as for instance structural biology, biophysics and molecular nanotechnology. In particular, revealing the distinct conformations of an individual protein is of utmost importance. Here, we show the imaging of individual proteins and protein complexes by low-energy electron holography. Samples of individual proteins and protein complexes on ultraclean freestanding graphene were prepared by soft-landing electrospray ion beam deposition, which allows chemical- and conformational-specific selection and gentle deposition. Low-energy electrons do not induce radiation damage, which enables acquiring sub-nanometer resolution images of individual proteins (cytochrome C and bovine serum albumin) as well as of protein complexes (hemoglobin), which are not the result of an averaging process.

**Significance Statement**

We report a novel method to image and reveal structural details of proteins on a truly single molecule level. Low-energy electron holography is used to image individual proteins electrospray-deposited on freestanding graphene. In contrast to the current state-of-the-art in structural biology, we do away with the need for averaging over many molecules. This is crucial since proteins are flexible objects that can assume distinct conformations often associated with different functions. Proteins are also the targets of almost all the nowadays known and available drugs. The design of new and more effective drugs relies on the knowledge of the targeted proteins structure in all its biologically significant conformations at the best possible resolution.

**Introduction**

Most of the currently available information on structures of macromolecules and proteins has been obtained from either X-ray crystallography experiments or cryo-electron microscopy investigations by means of averaging over many molecules assembled into a crystal or over a large ensemble selected from low signal-to-noise ratio electron micrographs, respectively (1). Despite the impressive coverage of the proteome by the available data, a strong desire for acquiring structural information from just one individual molecule is emerging. The biological relevance of a protein lies in its structural dynamics, which is accompanied by distinct conformations. For a protein to fulfil its vital functions in living organism, it cannot exist in just one single and fixed structure, but needs to be able to assume different conformations to carry out specific functions. Conceptually, at least two different conformations, just like in a simple switch, are needed. In view of oxygen transport to cells for example, binding oxygen in one specific conformation and releasing it again in a different conformation is needed. To address the "physics of proteins" as described by Hans Frauenfelder in his pioneering review (2), one needs to realize that proteins are complex systems assuming different conformations and exhibiting a rich free-energy landscape. The associated structural details however, remain undiscovered when averaging is involved. Moreover, a large subset of the entirety of proteins, in particular from the important category of membrane proteins, are extremely difficult, if not impossible, to obtain in a crystalline form. If just one individual protein or protein complex can be analyzed in sufficient detail, those objects will finally become accessible.

For a meaningful contribution to structural biology, a tool for single molecule imaging must allow for observing an individual protein long enough to acquire a sufficient amount of data to reveal its structure without altering it. The strong inelastic scattering cross-section of high-energy electrons as employed in the state-of-the-art aberration-corrected Transmission Electron Microscopes (TEMs) inhibits accumulation of sufficient elastic scattering events to allow high-resolution reconstruction of just one molecule before it is irremediably destroyed (3). The recent invention of direct detection cameras has dramatically pushed forward the effort in single particle imaging cryo-EM (4). In particular, it allows a reduction of the required number of objects for meaningful and high-resolution reconstructions. This technical innovation radically improves the signal-to-noise ratio for the same electron dose in comparison with a conventional charge-coupled-device camera coupled to a scintillator, a crucial aspect when imaging low atomic number and beam-sensitive material. Several research groups are trying to reduce the radiation damage problem by lowering the electron energy. To our knowledge, 20 keV is the lowest energy used in an aberration-corrected TEM (5). However, the radiation damage to biomolecules by electrons with a kinetic energy in the keV range will possibly never permit imaging of truly single proteins at atomic resolution (3, 6). Staining proteins with heavy metal atoms is unfortunately not a viable alternative, since it is well known that the chemical processes involved alter the protein structure (7). Moreover, heavy metal atoms are highly mobile under high-energy electron beams, which leads to ambiguous images. A recent approach to structural biology is associated with the X-ray Free Electron Laser (XFEL) projects. With this impressive technological development and novel experimental tool, it is now possible to elucidate the structure of proteins brought in the form of crystals of just nanometer size (8–11). This method even appeared as a way of gaining information at the atomic scale from just a single biomolecule. Meanwhile it has become clear that averaging over a large number of molecules will unfortunately not be avoidable (12). Future XFELs with orders of magnitude enhanced brightness and reduced pulse duration might eventually achieve the goal of single molecule imaging.

In contrast to the radiation damage problem experienced when using high-energy electrons or X-rays, biomolecules, for instance DNA, can withstand prolonged irradiation by electrons with a kinetic energy in the range of 50–250 eV. Even after hours of illumination and the exposure to a total dose of at least five orders of magnitude larger than the permissible dose in X-ray or high-energy electron imaging, biomolecules remain unperturbed as exemplified in a detailed study concerning DNA molecules (13). This, combined with the fact that the de Broglie wavelengths associated with this energy range are between 0.7 and 1.7 Å, makes low-energy electron microscopy techniques, especially holography, auspicious candidates for investigations at the truly single molecule level. In this lens-less microscopy scheme inspired by Gabor's original idea of holography (14), the samples are presented to a highly coherent beam of low-energy electrons generated by an atomically sharp field emitter (15–17) tip placed as close as 100 nm in front of the sample (Fig. S3). The interference pattern formed by the scattered and un-scattered electron waves, the so-called hologram, is recorded at an electron detector several centimeters away (for more details, see the Supplementary Information). Since in a hologram, the scattered and un-scattered electrons are contributing to the image formation, acquisition times as short as hundred microseconds are sufficient for high signal-to-noise ratio records (18). While highly coherent sources for low-energy electrons have been available for more than two decades, holography has long suffered from the lack of a substrate transparent to low-energy electrons but still robust enough that nanometer-sized objects can be deposited onto it. Recently, we have shown that ultraclean freestanding graphene fulfils these two requirements (19–21).

In the following, we show how sub-nanometer resolution images of individual proteins are obtained by means of low-energy electron holography. While the damage-free radiation of coherent low-energy electrons and the conceptual simplicity of the experimental scheme for holography are appealing, this tool for single protein imaging critically relies on the sample preparation method. The proteins must be brought into an ultra-high vacuum (UHV) environment and fixed in space for an appropriate period of time to accumulate sufficient structural information on the one hand, while avoiding the emergence of contaminants on the other hand. Here, native protein ions are transferred from aqueous solutions to the gas phase (22–27) and deposited onto ultraclean freestanding graphene in an UHV environment by means of soft-landing electrospray ion beam deposition (28–30) (ES-IBD). The workflow for imaging a single protein involves several steps, as illustrated in Fig. 1. An ultraclean freestanding graphene sample covering 500×500 nm$^2$ apertures milled in a 100 nm thick SiN membrane is prepared following the recently developed platinum metal catalysis method (31) and is characterized in the low-energy electron holographic microscope (Fig. 1, left). The sample is subsequently transferred to an ES-IBD system (Fig. 1, middle) under permanent UHV conditions by means of a UHV suitcase operating in the $10^{-11}$ mbar regime (see Supplementary Information for more details). Native cytochrome C (CytC), bovine serum albumin (BSA), and hemoglobin (HG) ion beams are generated by electrospray ionization and mass filtering. The charge states $z = 5–7$ are selected for CytC (22, 26) and the charge states $z = 15–18$ are selected for BSA (32). In the case of HG, the charge states $z = 16$ or $z = 17$ of the intact complex are known to be of native conformation (33) and hence the corresponding *m/z* region is selected (the corresponding mass spectra are displayed in the Supplementary Information). In all three cases, the ions are decelerated to a very low kinetic energy of 2–5 eV per charge, which ensures retention of the native state upon deposition onto ultraclean freestanding graphene (26). Preparative mass spectrometry of proteins (28, 34, 35) followed by electron microscopy has already been reported in the literature, but there the deposition was made on thick carbon films and not performed under UHV conditions (36, 37). Furthermore, the imaging by high-energy electrons in a TEM required exposure to atmospheric conditions and negative staining of the proteins.

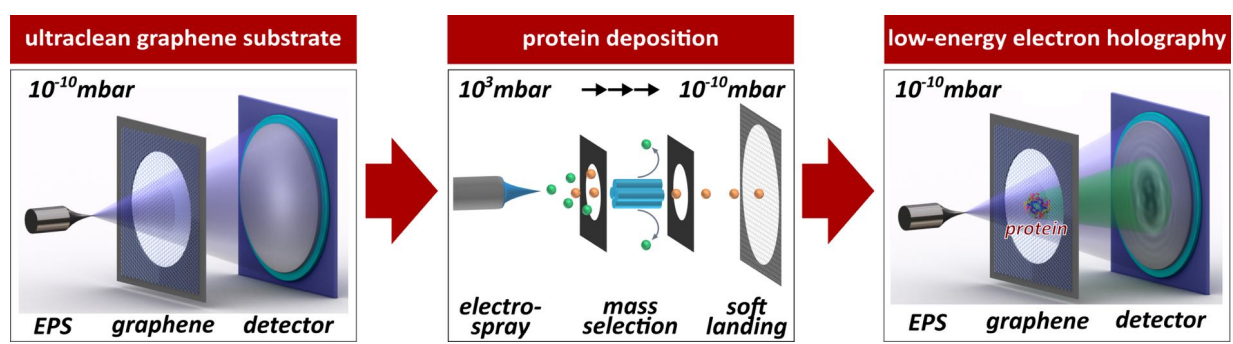

***Fig. 1. Schematic workflow for imaging a single protein****. From left to right: an ultraclean graphene sample is characterized in a low-energy electron holographic microscope. Deposition of proteins onto freestanding graphene in an m/z filtered ES-IBD system. Imaging of the proteins within the previously characterized region by means of low-energy electron holography. An electron point source (EPS) emits a divergent beam of highly coherent low-energy electrons and holograms of the deposited proteins are recorded on the detector. Throughout the experimental workflow, the sample is kept under strict UHV conditions with the help of an UHV suitcase for the transfer between the two experimental chambers (see Supplementary Information for more details).*

**Results**

**Cytochrome C.** After deposition, the samples are transferred again under preserved UHV conditions from the ES-IBD system back to the low-energy electron holographic microscope, where holograms of individual proteins are recorded. In Fig. 2, holograms of three distinct entities found on freestanding graphene after deposition of native CytC ions are presented [Fig. 2(a)–2(c)]. In these images, the characteristic fringes of a holographic record, resulting from the interference between the un-scattered and elastically scattered electrons, are observed. From the hologram itself it is impossible to recognize the shape of the object. Numerical hologram reconstruction involving back propagation of the wave front from the hologram to the sample plane (38–40) is required to finally reveal the object's structure [Fig. 2(d)–2(f)]. For instance, in Fig. 2(d), an individual CytC is displayed after hologram reconstruction. The diffuse rings around the object are due to the presence of the out-of-focus twin image inherent to in-line holography (38). As apparent from the high contrast images shown in Fig. 2(d)–2(f), not only are the globular structures with the correct overall dimensions of the protein revealed, but also details of the shape of CytC in different orientations even while forming agglomerates of two (Fig. 2e) respectively three proteins (Fig. 2f). The spatial resolution attained in a hologram can be estimated using the Abbe criterion (41, 42) and by measuring the largest angle under which interference fringes are observable (40, 43, 44). In Fig. 2, a resolution of 7–8 Å is calculated and the same value is found in the reconstructed images by measuring the edge response (45) over the protein structures. A similar resolution is estimated for all other micrographs presented below. In a hologram, the spacing between consecutive interference fringes gradually decreases towards higher orders. Hence, high-order interference fringes and consequently high-resolution structural details are most susceptible to mechanical vibrations. The latter currently limit the resolution, and intense efforts are ongoing to increase the mechanical stability of the low-energy electron holographic microscope in order to overcome this limitation and approach atomic resolution. While the current resolution already reveals the outer shape of single proteins and protein sub-units, an enhanced resolution of 2 Å will permit imaging internal structural details as well. The inner contrast variations apparent in the images presented here demonstrate that proteins are sufficiently transparent to low-energy electrons [Fig. 2(d)–2(f)].

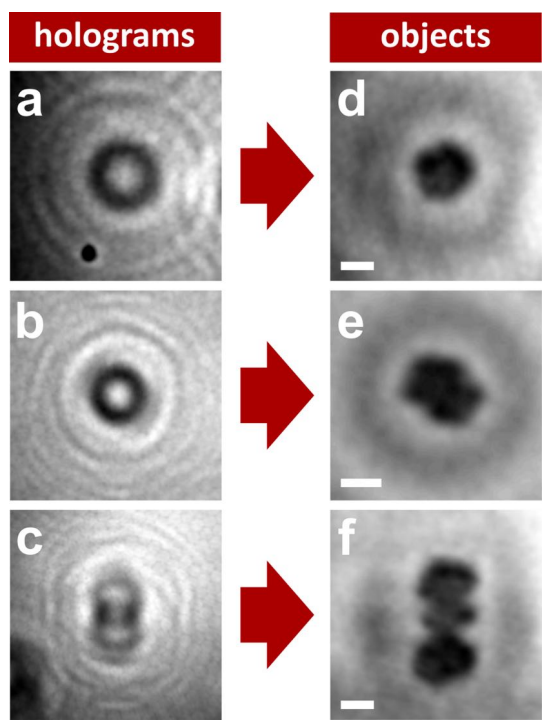

***Fig. 2. Low-energy electron holograms of cytochrome C and their reconstructions**. (a)–(c) Three holograms of CytC. **(d)–(f)** Numerical reconstructions revealing the shapes of the objects. The diffuse rings around the object are due to the presence of the out-of-focus twin image inherent to in-line holography. The scale bars correspond to 2 nm.*

A very important question for the imaging of soft-landed proteins by means of low-energy electron holography is the charge state of the adsorbed particles after landing. While several studies have shown retention of the charges states for protein deposited onto self-assembled monolayer (46–48) other studies have observed neutral species after landing on metallic surfaces (49, 50). It has recently been shown that low-energy electron holography is sensitive to electric charges as small as a fraction of elementary charge (51). The presence of a charge acts like a lens bending the electron trajectories around the object and thus preventing the reconstruction of the object's shape. In the protein images presented here there is no evidence for charged objects, we therefore conclude that the protein ions, produced during the electrospray process, are neutralized after landing on the graphene substrate.

In Fig. 3, a complete data set corresponding to the experimental workflow described above (Fig. 1) is presented for the case of soft-landing and imaging of CytC. In this figure, an area of ultraclean freestanding graphene prior to deposition is characterized by means of low-energy electron holography and exhibits very few contaminations on the otherwise homogenous and clean surface [Fig. 3(a)]. A mass-selected beam of native CytC ions is prepared, which shows the characteristically low charge states corresponding to the protein in its folded state (22) [see Fig. 3(b)]. In Fig. 3(c), a survey image of the very same freestanding graphene region after a complete transfer and deposition cycle is displayed. After the deposition of a fraction of a monolayer of proteins, well-separated globular objects of similar size as well as agglomerations thereof, probably as a result of surface diffusion, are found on the graphene substrate. Control experiments involving the complete transfer process between the two vacuum chambers but without deposition have been performed, demonstrating that this process does not introduce any contaminants onto freestanding graphene (see Fig. S4). At high magnification, the shape of the individual CytC proteins is revealed in several distinct orientations on graphene [Fig. 3(d)–3(j)]. It is not surprising to find the protein in different orientations since the deposition process is random in this respect.

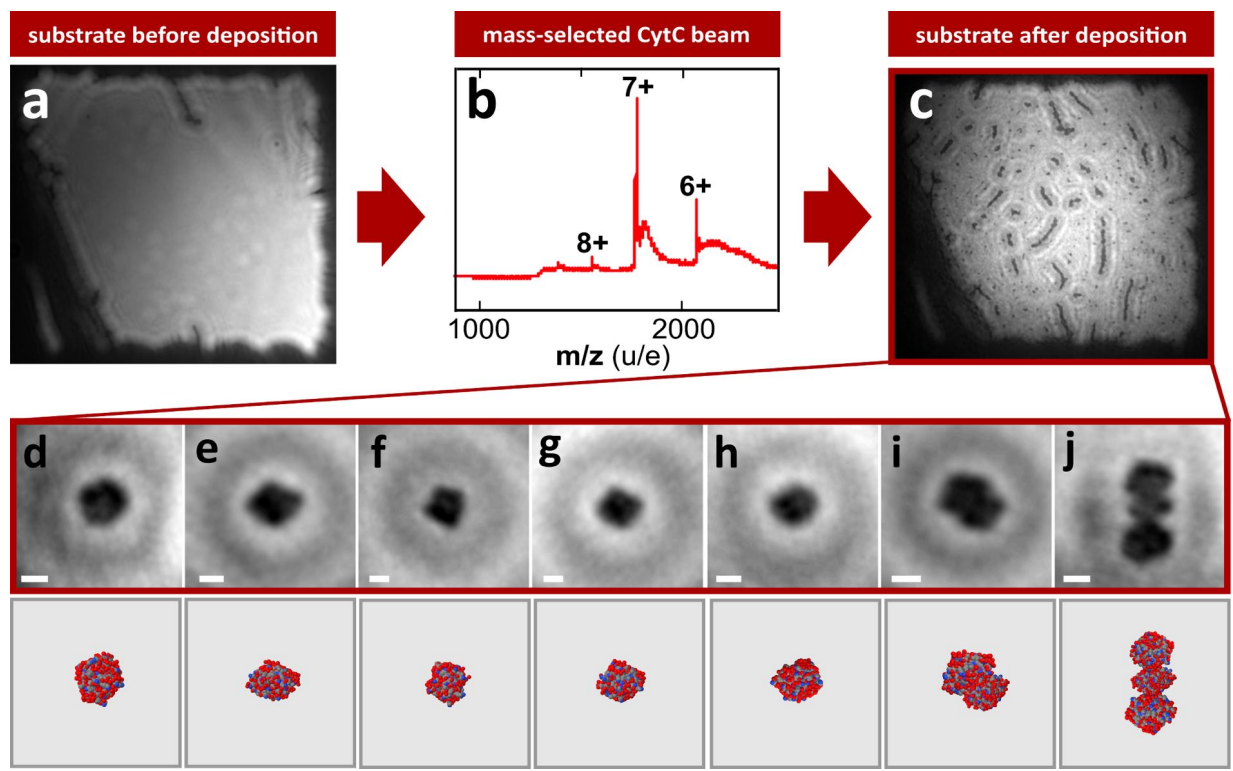

***Fig. 3. Complete data set for the imaging of CytC***. *(a) Low-energy electron image of ultraclean graphene covering an 500×500 nm$^2$ aperture before protein deposition. (b) Mass-spectrum of the mass-selected CytC beam. (c) A survey image of the very same freestanding graphene region after deposition of CytC. (d)–(j) Low-energy electron micrographs with suggestions for possible protein orientations based on the averaged protein structure derived from X-ray crystallography data and documented in the protein data bank (pdb id: 1HRC). The scale bars correspond to 2 nm.*

The high-magnification low-energy electron micrographs of CytC presented in Fig. 3 are of sufficiently high resolution to allow comparison with the structural data information obtained from X-ray crystallography investigations and available from the protein data bank (pdb id: 1HRC). The overall size of the imaged CytC corresponds to the expected dimensions, and the low-energy electron images can clearly be associated with proteins in several distinct orientations. Imaging single objects over an extended period of time never led to any changes in the images. In particular, no sign of decomposition of the protein during electron exposure was observed, similar to what was demonstrated previously with DNA (13).

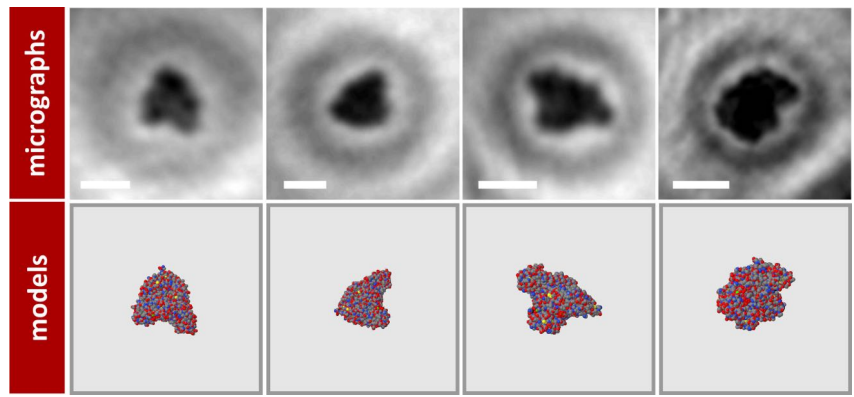

*Fig. 4. Low-energy electron micrographs of BSA in different orientations on graphene. **Top,** low-energy electron micrographs of BSA. **Bottom,** the atomic model of BSA (pdb id: 3V03) in the corresponding orientations. The scale bars correspond to 5 nm.*

**Bovine Serum Albumin.** The same experimental workflow was used in the case of imaging BSA, a much larger protein than CytC (66 kDa versus 12 kDa). A collection of low-energy electron micrographs of BSA is presented in Fig. 4 (top). Similar to CytC, high-contrast images reveal features that suggest a globular structure with the correct dimensions of the protein. In contrast to CytC, which is nearly spherical in shape, the three-dimensional shape of BSA is traditionally described as heart-shaped. The micrographs of individual BSA molecules reflect this structure as well as the protein in other but very characteristic orientations. The agreement between the micrographs and the atomic model for a protein like BSA, not purely globular but exhibiting very pronounced structural features, clearly demonstrates that proteins can be found in UHV in structures closely related to their native structure.

A further possibility to qualitatively analyze our micrographs is to compare them with simulated electron density maps at the corresponding resolution. In Fig. 5, two of the BSA micrographs presented in Fig. 4 are therefore compared to electron density maps simulated at a resolution of 8Å with the software *Chimera* (52), originally developed for the analysis of Cryo-EM images. We find a considerable correspondence between the simulated density maps and the micrographs and the resolution estimate made from the holographic record is in good agreement with the simulation. To demonstrate that even proteins as large as BSA are sufficiently transparent for low-energy electrons for delivering three-dimensional structural information from a single hologram, the contrast in these micrographs has now been enhanced in comparison to Fig. 4 by using an image processing software while maintaining a linear look-up table but only changing the minimum and maximum grey-level. Especially in the low-energy electron image presented in the right column, variations of the inner contrast can easily be recognized. To understand the origin of the darker region in the center of the protein image and to associate it with structural features, one may take a look at the protein from a direction parallel to the plane of graphene. The red arrows in the middle row of Fig. 5 indicate the direction of observation corresponding to the side views presented in the third row of this image. With the help of these side view representations of the density maps, in which graphene is indicated by a solid blue line, it is possible to address how the proteins are actually adsorbed on the graphene substrate. Of particular interest is the side view presented in the right column as it shows that in this specific orientation, BSA is of strongly varying thickness with a maximum towards the center. The darker region observed in the micrograph evidently corresponds to a higher absorption due to an increased protein *thickness* in this area. This observation and analysis demonstrate that already with a resolution of 8Å, information on the three-dimensional structure of the protein can be gained. In prospect of an improved resolution of the order of 1-2Å, it also illustrates the future ability to gain a complete three-dimensional structure from a single low-energy electron hologram of proteins at least as large as 60kDa.

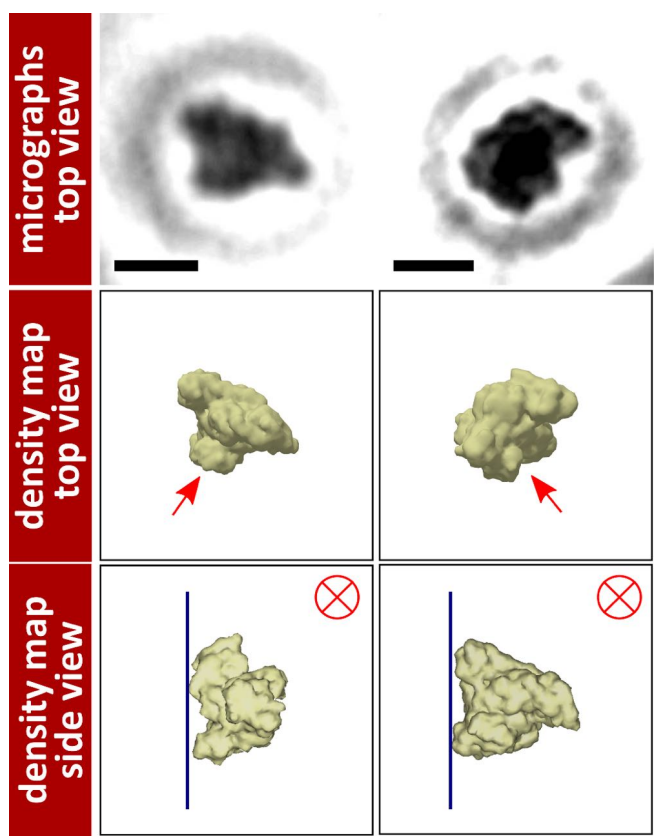

*Figure 5: Comparison of low-energy electron micrographs of BSA with simulated electron density maps*. **Top,** *low-energy electron micrographs of BSA. The scale bars correspond to 5 nm.* **Middle,** *electron density maps simulated at a resolution of 8Å and rotated to match the orientation of the proteins presented in the top row.* **Bottom,** *side view of the density map along the directions of observation indicated by the arrows shown in the middle row.*

In biology, a non-covalently bonded complex of several proteins rather than a single protein is performing a function. Next to the atomic structure of a protein, the composition and structure of protein complexes are of utmost importance. The data of Fig. 3 show that protein agglomerations formed of two and three CytC can be resolved. It has been extensively shown that by means of electrospray ionization, it is possible to ionize entire protein complexes while keeping their native conformations (24, 53–55).

**Hemoglobin.** In Fig. 6, two micrographs of individual hemoglobin, a complex of four protein subunits, are presented, demonstrating that with our method, entire protein complexes in their native configuration can be deposited and individually imaged. While for the cases of CytC and BSA the agreement between the low-energy electron images and the atomic models is almost perfect, differences can be observed for the case of HG. Since HG is a protein complex composed of four sub-units, it exhibits a large conformational flexibility, actually required for its function in a living organism. When an averaging process over millions of molecules is involved in the imaging of a highly flexible protein, discrete conformations cannot be distinguished and only an average structure evolves. However, with a technology capable of imaging individual proteins, like low-energy electron holography, the entire conformational landscape is revealed. It is therefore not surprising that structural differences between the low-energy electron micrographs and the atomic model are apparent. Furthermore, a much larger set of images will be needed to address the full conformational landscape of flexible proteins, as for instance hemoglobin, ~~and~~ to be able to determine a protein structure, and judge the abundance of the variants. Here, we present two images of hemoglobin that could be associated to its atomic model in a specific orientation.

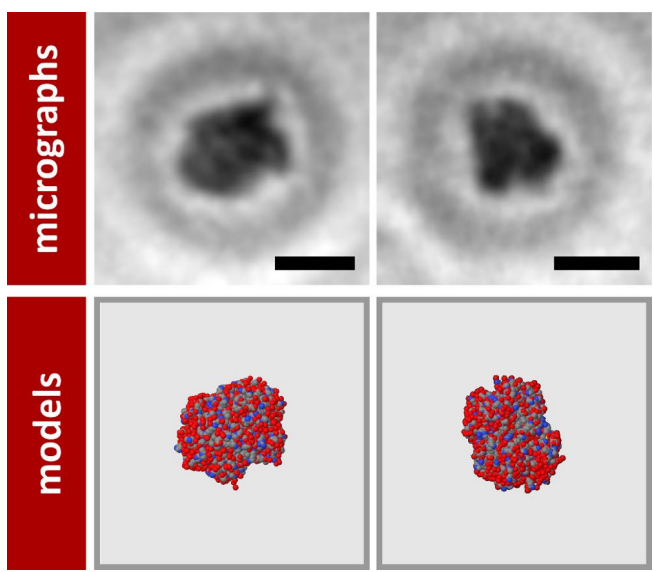

*Fig. 6. Low-energy electron micrographs of two individual hemoglobin molecules and the atomic model in the corresponding orientations. **Top,** two micrographs of hemoglobin soft-landed onto freestanding graphene. **Bottom,** suggestions for possible orientations based on the averaged protein structure derived from X-ray crystallography data and documented in the protein data bank (pdb id: 2QSS). The scale bars correspond to 5 nm.*

**Conclusion and Outlook**

The ultimate goal of directly uncovering the structure of unknown proteins or protein complexes and describing their conformations at the atomic level still requires experimental efforts towards a better chemical and conformational selectivity of the deposition process. This shall in the future be attained by adding ion-mobility capability to the ES-IBD device. With this more elaborated deposition device, it will be possible to select the objects not only on the basis of their charge state but also of their gas phase conformation. By this, an assessment of the relation between the conformation in the gas-phase and on the surface, influenced by the charge state, (22–25) deposition conditions (49, 56) and surface properties will be possible.

The mapping of the proteins with unknown structure will also require an improved imaging resolution, fundamentally limited solely by the electron wavelength. Furthermore, as three-dimensional information is encoded in a single in-line hologram, improved spatial resolution will already permit to determine the (x,y,z) spatial coordinates of every atom of a protein from this very holographic record. A complementary strategy to reveal the complete three-dimensional structure of a single protein is to add tomographic capability to the experimental setup. At this stage, the comparison of the low-energy electron micrographs with atomic models available at the protein data bank has the character of a control experiment, proving the feasibility of this novel methodology.

Nevertheless, fundamental questions remain to be addressed. Most crucial is the influence of the environment on the protein's structure. For proteins the deposition as native gas phase ions onto graphene in vacuum where they are neutralized definitely represents a significantly different environment than the aqueous medium of the cell. There is a significant body of strong evidence, especially from ion-mobility/mass-spectrometry investigations, demonstrating that proteins and protein complexes can be transferred from the liquid phase to a vacuum environment while maintaining their tertiary, respectively quaternary, structures unperturbed (25, 57–60). The low-energy electron micrographs presented here are further strong evidences that proteins in a folded state are stable in UHV. As recently demonstrated it is possible to add water molecules to small peptides (61) directly in an UHV environment. Low-energy electron holography with its ability to image proteins individually will also allow to study the effects of adding hydration shells to the protein. Furthermore, questions related to transport, such as diffusion of proteins and subsequent association into protein complexes, will be addressed. First observations of the diffusion of folded proteins on freestanding graphene by means of low-energy electron holography are presented in Fig. S5, showing the potential of the method described here for accessing dynamic processes as well.

To conclude, we have shown here how to image a single protein by combining ES-IBD technology with low-energy electron holography. This has led to the first ever tool for revealing structural details of single native proteins and protein complexes without destroying them. With the recent advances in electrospray ionization and mass spectrometry of large protein complexes (62), and in particular membrane proteins (53, 55), even the structure of these biologically important but reluctant to readily crystalize entities may become accessible in the future.

**Materials and Methods**

Ultraclean freestanding graphene is prepared by the Pt-metal catalysis method described in detail elsewhere (31). Prior to the transfer of the ultraclean substrate from the UHV chamber of the low-energy electron holographic microscope to the UHV chamber of the ES-IBD device, the cleanliness of the substrate is characterized and reference images are recorded for comparing the very same region of freestanding graphene before and after protein deposition.

During the whole experimental workflow, the samples are kept under strict UHV conditions with the help of a UHV suitcase for transfer between the two experimental chambers. Details of the ES-IBD procedure and of the low-energy electron holography experimental scheme are described in the supplementary information document.

**Acknowledgment**

We thank Frank Sobott and Ester Martin for advice on preparing native protein beams. The work presented here is financially supported by the Swiss National Science Foundation (SNF). We also appreciate support by the EU commission for building the equipment in the frame of the former SIBMAR project.

**Author Contributions**

J.-N. L. had the original idea to combine ES-IBD and low-energy electron holography and further elaborated the concept with K.K and H.-W.F. J.-N. L. prepared the ultraclean freestanding graphene supports and recorded the holograms. S.R. and J.-N. L. planned the deposition experiments and along with S.A. performed the electrospray deposition of the proteins onto graphene. T.L. performed the hologram reconstructions with her self-developed software package. S.R. and J.-N. L. interpreted the data. H.-W. F. invented the technology of lens less holography with low-energy electrons based on atomic sized coherent electron point sources. C.E., T.L., J.-N. L., & H.-W. F. further developed the low-energy electron holographic microscope used in this study. S.R. and K.K. developed the ES-IBD technique. J.-N. L, C. E. & H.-W. F. wrote the manuscript main text and with S.R the supplementary information, in discussions with all remaining authors.

**Supplementary Information**

**1. ES-IBD**

Soft-landing electrospray ion beam deposition takes place in a home-built instrument (29, 30) (Fig. S1). The ion beam is generated by a nano electrospray source with an optimized hydrodynamic vacuum transfer (63) at a flow rate of 20-30 µL/h and an emitter voltage of approximately 3 kV. The positive ions enter the vacuum through a heated metal capillary and are collimated in an ion funnel and a collisional collimation quadrupole operated in *rf*-only mode. In the third pumping stage a mass filtering quadrupole selects the m/z region of interest, which is monitored by the time-of-flight mass spectrometer in the fourth pumping stage. Here, a retarding field energy analyser measures the kinetic energy of the ions. The beam is then guided by electrostatic lenses towards the target in the 6$^{th}$ pumping stage being at $2\times10^{-10}$ mbar, where the protein deposition takes place. To ensure gentle landing, the collision energy is controlled by applying a bias voltage to the target and hence reducing the kinetic energy of the protein ions to 2-5 eV per charge.

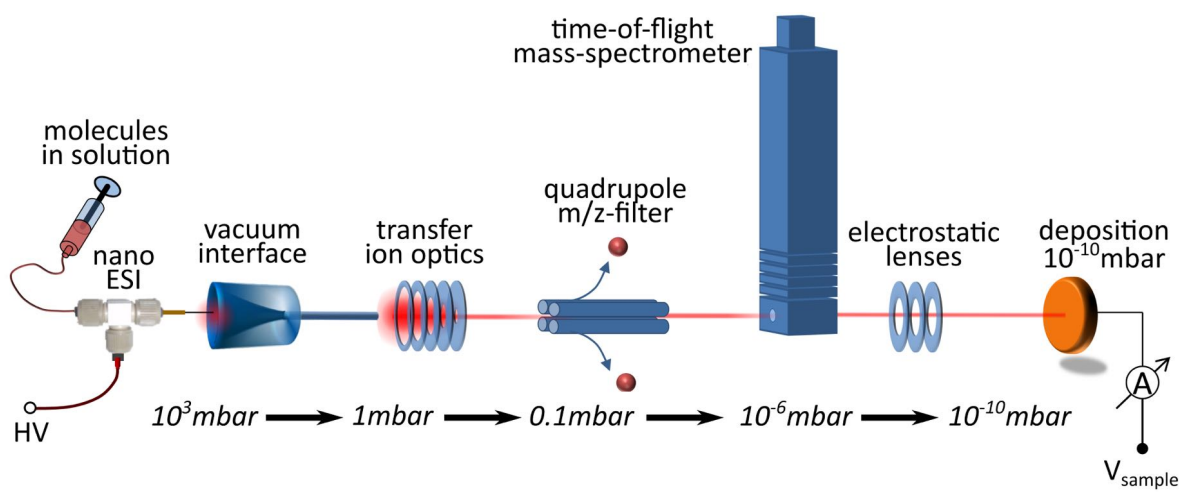

***Fig. S1.*** *Illustration of the working principle of the home-built instrument for soft-landing electrospray ion beam deposition of proteins. The proteins are directly deposited from the liquid phase onto a substrate kept at UHV conditions in order to avoid contaminations.*

To obtain ion beams of native CytC (bovine, Fluka 30398), a solution of 0.15 mg/mL was prepared in aqueous 50 mM ammonium acetate buffer. With a spray flow rate of 25 µL/hr, an ion current of 1.1 nA is detected at the TOF-MS. Unfiltered mass spectra (Fig. S2 left/top) show low charge states of +5 to +7, corresponding to folded CytC. At lower m/z values peaks corresponding to highly charged (z>+8) unfolded CytC and peaks that relate to fragments or contamination are found. Note that, due to the limited dynamic range of the TOF-MS, the detector amplification was set very high, such that the peaks of the native CytC are distorted. The m/z selective quadrupole was tuned to select a m/z-window from 1250 u/e to approximately 3500 u/e by setting a *rf*-amplitude of 700 V with a differential dc-voltage of 5%. This results in a beam of predominantly native CytC ions (Fig. S2 left/bottom) from which unfolded proteins (low m/z) and undefined heavy aggregates (high m/z) are removed. The current of the m/z filtered ion beam amounts to 20 pA at the substrate. A total charge of approximately 20 pAh is deposited while the pressure is maintained in the $10^{-10}$ mbar range.

Ion beams of native bovine serum albumin (BSA) are generated following the same general procedure as for CytC. Solutions are prepared by dissolving BSA in aqueous 50mM ammonium acetate buffer at a concentration of approximately $10^{-5}$ M. The unfiltered mass spectrum displays intense peaks around charge state z=16, which are related to native BSA (32), whereas higher charge states can be expected to be at least partially unfolded. Further, at m/z>6000 u/e unresolved intensity may be related to unspecific agglomeration. Therefore, the region between 3250 u/e and 6000 u/e is selected for deposition. The current of the m/z filtered beam amounts to 8 pA at the substrate. A total charge of approximately 20 pAh is deposited while the pressure is maintained in the $10^{-10}$ mbar range.

Hemoglobin (bovine, Sigma H2500) is a protein complex of four myoglobin subunits, two A and two B. Ion beams are generated from solutions of 0.3 mg/mL HG prepared in 50mM ammonium acetate buffer. A current of 600pA is detected at the TOF-MS. The mass spectrum is very complex and is resolved only partially due to the limited performance concerning resolution and dynamic range of the home-built linear TOF-MS. Nevertheless, a comparison with literature spectra allows identifying characteristic peaks (Fig. S2 right/bottom), which become more pronounced after removal of the unspecific agglomeration in the high m/z range (>5000 u/e) by mass filtering with the quadrupole. The beam transmitted for deposition contains native intact HG complexes ($Q^{17+}$, $Q^{16+}$) along with dimers (D) and monomers (HBA). A total charge of 70 pAh was deposited

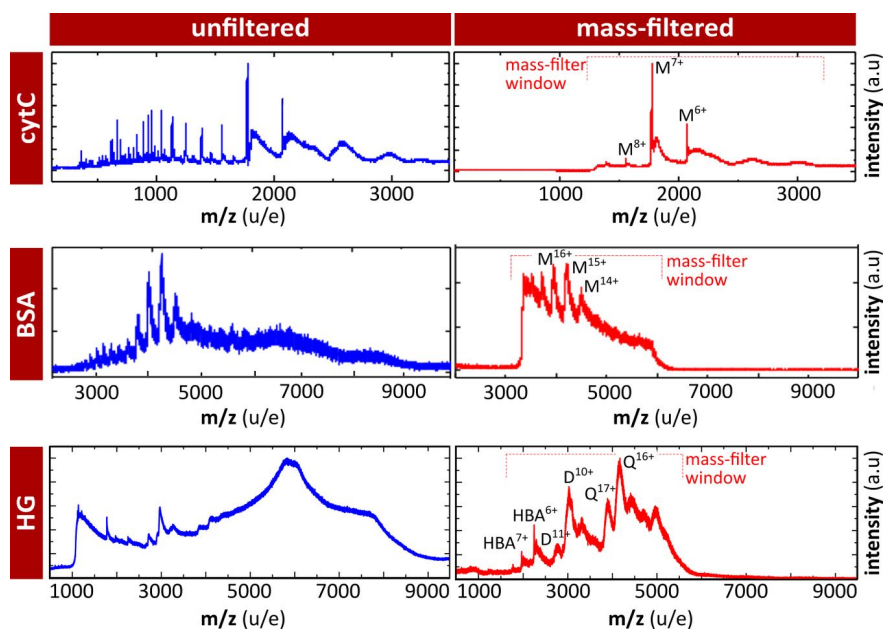

***Fig. S2.*** *Mass spectra of Cytochrome C (top), BSA (middle) and Hemoglobin (bottom). Left: the m/z spectra before mass filtering are displayed. Right: the corresponding mass-filtered spectra.*

## 2. Low-Energy Electron Holography

In the low-energy electron holographic setup (15), inspired by Dennis Gabor's original idea of in-line holography (14, 64), a sharp (111)-oriented tungsten tip acts as electron point source (EPS) providing a divergent beam of highly coherent electrons (65, 66). The atomic sized electron field emitter can be brought as close as 100nm to the sample with the help of a 3-axis nanopositioner. Part of the electron wave is elastically scattered off the object and hence is called the object wave, while the un-scattered part of the wave represents the reference wave. At a distant detector, the hologram, i.e. the pattern resulting from the interference of these two wave fronts is recorded. The detector unit comprises a multi-channel plate (MCP) for signal enhancement, a phosphor screen for translating the electron signal in an optical image, a fibre optic plate for guiding this image out of the vacuum system and a CCD camera for recording. In order to discriminate secondary electrons, a small retarding potential can be applied to the entrance of the MCP. This is particularly necessary when imaging metallic clusters or metallic thin films producing a lot of secondary electrons when exposed to the low-energy electron beam. Our experience is that when imaging low atomic number materials as for instance carbon fibres or biological materials no substantial signal originating from the production of secondary electrons at those materials can be detected and the discriminator voltage at the MCP is obsolete. These observations confirm also that predominantly elastic scattering occurs at biological objects like proteins.

The magnification of the imaging system is given by the ratio between detector-to-source distance $Z$ and sample-to-source distance $z$. Given that in the present system $Z$=70mm and that $z$=100nm-3mm, the magnification can be as high as $10^5$. The electron energy is dictated by the sample-to-source distance and the sharpness of the field emitter tip for a given electron emission current. All holograms presented here were recorded with electron energies between 60eV and 140eV and an emission current of typically 50nA, which corresponds at high magnification to a current density of 5pA/nm$^2$ or $10^7$ electrons/s·nm$^2$ for a field of view of 100x100nm$^2$. A typical acquisition time for a hologram amounts to 100-200ms corresponding to a total number of emitted electrons of the order of $10^{10}$. A hologram, in contrast to a diffraction pattern, contains also the phase information of the object wave, and the object structure can thus be unambiguously reconstructed. After background subtraction, the shape of the objects is revealed by numerical reconstruction from the hologram, which is essentially achieved by back propagation to the object plane, which corresponds to evaluating the Fresnel-Kirchhoff integral transformation. The individual steps related to the numerical hologram reconstruction are listed in section 4 below. A detailed description of the reconstruction procedure can be found elsewhere (38–40). In order to enable an artefact-free reconstruction, the proteins are deposited on graphene, which is known to be highly transparent for low-energy electrons (20). To attain such high transparency, the graphene samples were prepared by the Pt-metal catalysis method described elsewhere (31).

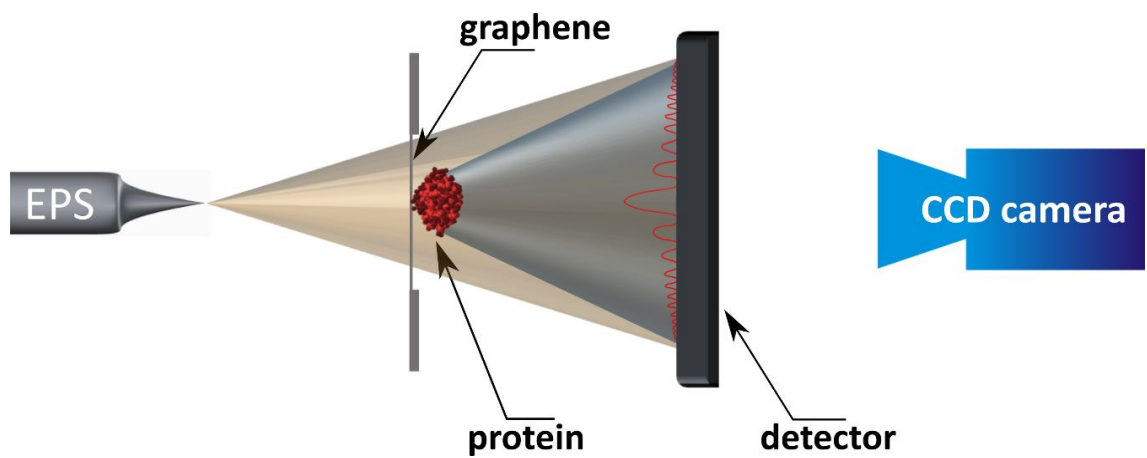

*Fig. S3*. Low-Energy Electron Holography Scheme

### 3. UHV Transfer

A vacuum suitcase, Fig. S4(a) (*Ferrovac GmbH, Zurich*), is used to transfer samples between the two UHV based experiments, electron holography on the one side, and electrospray ion beam deposition on the other side.

The suitcase is equipped with a SAES getter/ion getter pump combination operated by a battery driven power supply. The pressure is kept below $2\times10^{-10}$ mbar at all times, except for the short duration of the transfer (1-2 minutes) where it rises into the $1\times10^{-9}$ mbar regime. The performance of this suitcase, originally developed for STM experiments is well known from various surface science studies (67, 68).

Each instrument is equipped with a load-lock to which the suitcase can be attached. The load-lock is pumped by a turbo molecular pump supported by a cryogenic active charcoal trap at $LN_2$ temperature. A pressure in the $10^{-9}$ mbar range is established within a few hours, ensuring a contamination-free transfer of the samples. Fig. S4 shows low-energy electron projection images of the very same freestanding graphene region before (b) and after (c) the transfer from the low-energy electron holographic microscope located in Zurich to the ES-IBD chamber in Stuttgart and back to the microscope in Zurich. No relevant sign of contamination due to transfer and transport is observed.

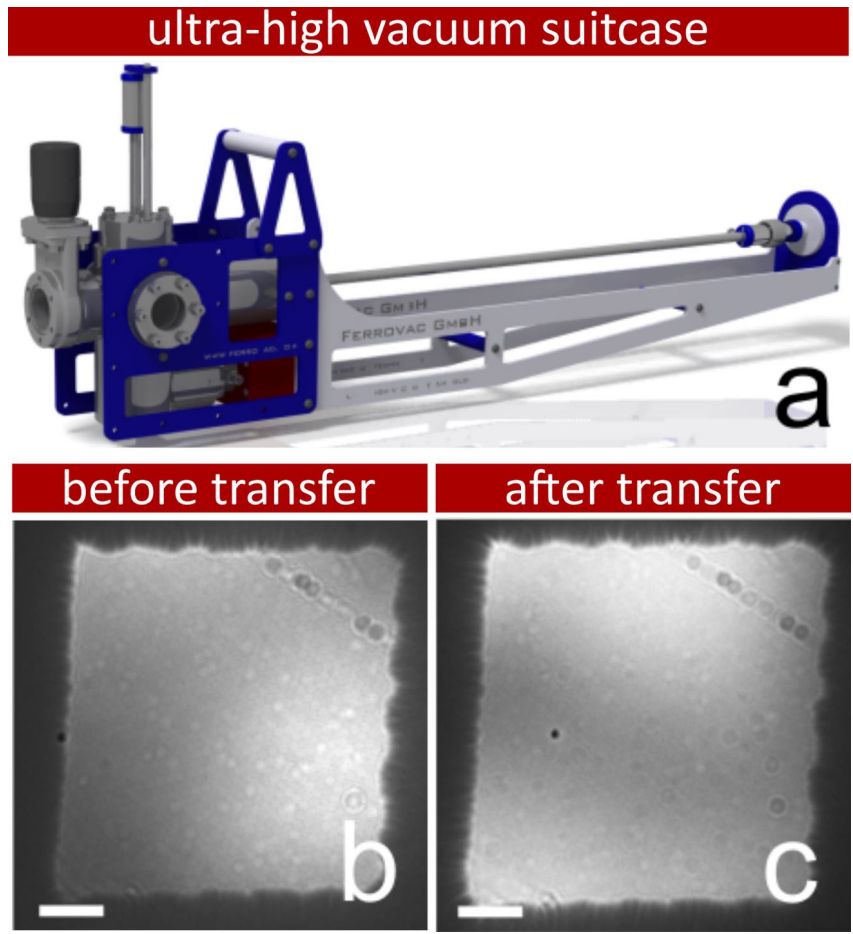

***Fig. S4.*** *UHV Vacuum Suitcase and its performance.* ***a,*** *Three-dimensional rendering of the transport suitcase enabling UHV transfer of ultraclean freestanding graphene between the low-energy electron holographic microscope and the ES-IBD chamber. Low-energy electron projection images before (**b**) and after (**c**) a complete transfer and travel cycle without protein deposition. In the course of the transfer procedure between the two vacuum chambers no relevant contamination of the graphene has built up. Scale bar accounts for 100nm.*

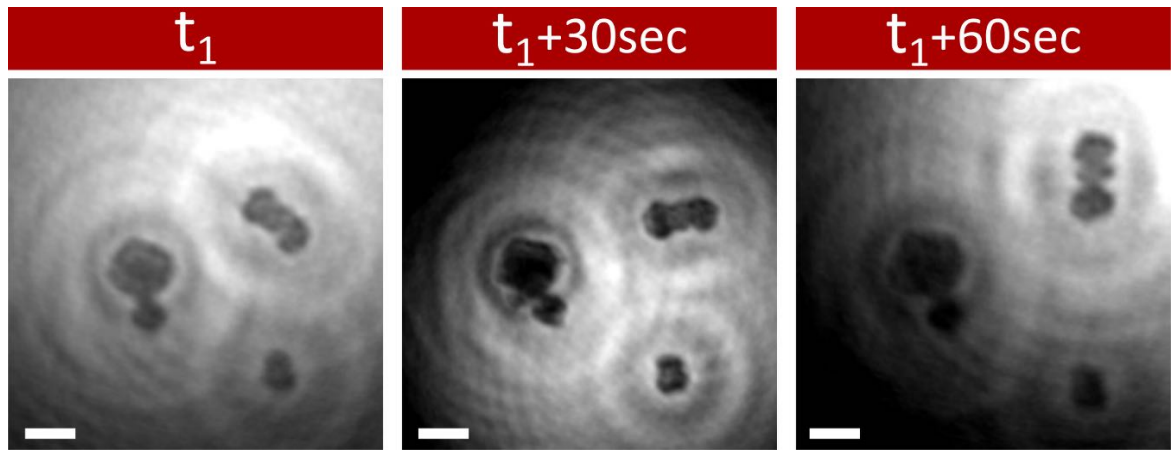

***Fig. S5.*** *Time evolution of the orientation of CytC complexes. The time lapse between subsequent observations amounts to 30sec. From these images it is evident that at least some of the deposited proteins are mobile on freestanding graphene. Thus, low-energy electron holography appears to be a method for also studying diffusion of proteins on surfaces. This observation suggests that a low-energy electron holographic microscope operating at cryogenic temperatures might be needed in order to fix the protein in space and attain atomic resolution. The scale bars correspond to 5nm.*

## 4. Numerical reconstruction of holograms

The numerical recipes for the reconstruction of digital holograms are described in detail elsewhere(40, 69). The reconstruction code based on the algorithm presented in (40) is available for download on the MATLAB File exchange server (File ID: #59940).

The numerical reconstruction of digital holograms consists of two steps. 1. The first step involves the multiplication of the hologram distribution $H(x_H, y_H)$ with the simulated complex-valued reference wave distribution $R = e^{ikr_H} / r_H$

$$U_H(x_H, y_H) = R(x_H, y_H) H(x_H, y_H) \tag{1}$$

where $r_H = (x_H, y_H, z_H)$ denotes the coordinates in the hologram plane and $z_H$ is the source-to-detector distance. 2. The second step involves the back-propagation of the complex-valued wavefront $U_H(x_H, y_H)$ distribution from the hologram plane to the object plane, whereby the propagation is described by the integral transform based on the Huygens-Fresnel principle:

$$U(r_O) \approx -\frac{i}{\lambda z} \iint U(r_H) \frac{\exp(-ik|r_H - r_O|)}{|r_H - r_O|} d\sigma_H, \tag{2}$$

with $r_O = (x_O, y_O, z_O)$ being the coordinate in the object plane. Details concerning evaluating the integral transform given by Eq. 2 are described elsewhere (40, 69).